\documentclass[a4paper,11pt]{article}

\usepackage[a4paper,left=80pt,right=75pt,top=80pt,bottom=100pt]{geometry}
\usepackage[latin1]{inputenc}
\usepackage{dsfont}
\usepackage{amsfonts,amsmath,amssymb}
\usepackage{amsthm}
\usepackage{graphicx}
\usepackage[small,bf]{caption}
\usepackage{subcaption}
\usepackage{booktabs}
\usepackage[numbers, sort]{natbib}
\allowdisplaybreaks[1]
\usepackage{color}
\usepackage{ulem}
\usepackage{tikz}
\usepackage{multirow}
\usetikzlibrary{arrows}
\usepackage{float}

\usepackage[pagebackref=false]{hyperref}

\def\bal#1\eal{\begin{align}#1\end{align}}
\newcommand{\be}{\begin{equation}}
\newcommand{\ee}{\end{equation}}
\newcommand{\bea}{\begin{eqnarray}}
\newcommand{\eea}{\end{eqnarray}}
\newcommand{\besub}{\begin{subequations}}
\newcommand{\eesub}{\end{subequations}}
\newcommand{\ba}{\begin{array}}
\newcommand{\ea}{\end{array}}
\newcommand{\bi}{\begin{itemize}}
\newcommand{\ei}{\end{itemize}}

\begin{document}

\begin{titlepage}

\begin{center}
{\LARGE
SUGRA New Inflation with Heisenberg Symmetry
}
\\[15mm]

Stefan Antusch$^{\star\dagger}$,~
Francesco Cefal\`a$^{\star}$
\end{center}
\addtocounter{footnote}{0}

\vspace*{0.20cm}

\centerline{$^{\star}$ \it
Department of Physics, University of Basel,}
\centerline{\it
Klingelbergstr.\ 82, CH-4056 Basel, Switzerland}

\vspace*{0.4cm}

\centerline{$^{\dagger}$ \it
Max-Planck-Institut f\"ur Physik (Werner-Heisenberg-Institut),}
\centerline{\it
F\"ohringer Ring 6, D-80805 M\"unchen, Germany}

\vspace*{1.2cm}

\begin{abstract}
We propose a realisation of ``new inflation'' in supergravity (SUGRA), where the flatness of the inflaton potential is protected by a Heisenberg symmetry. Inflation can be associated with a particle physics phase transition, with the inflaton being a (D-flat) direction of Higgs fields which break some symmetry at high energies, e.g.\ of GUT Higgs fields or of Higgs fields for flavour symmetry breaking. This is possible since compared to a shift symmetry, which is usually used to protect a flat inflaton potential, the Heisenberg symmetry is compatible with a (gauge) non-singlet inflaton field. In contrast to conventional new inflation models in SUGRA, where the predictions depend on unknown parameters of the K\"ahler potential, the model with Heisenberg symmetry makes discrete predictions for the primordial perturbation parameters which depend only on the order $n$ at which the inflaton appears in the effective superpotential. The predictions for the spectral index $n_s$ can be close to the best-fit value of the latest \textit{Planck 2013} results. 
\end{abstract}

\end{titlepage}
\newpage

\section{Introduction}

The recent CMB results of the \textit{Planck} satellite \cite{Planck} have provided further strong support for the inflationary paradigm \cite{inf1,inf2,inf3}, pointing at a Gaussian adiabatic spectrum of primordial perturbations which is almost, but not exactly, scale invariant. The spectral index $n_s$ has been measured to be $n_s = 0.9603 \pm 0.0073$ \cite{Planck}, which already provides a good discriminator between inflationary models \cite{inf4}. 
In about one year, the polarisation data of \textit{Planck} will be released, and the results for primordial gravitational waves, expressed in the tensor-to-scalar ratio $r$, will be presented. This will be a further discriminator between different types of inflation models. For example, simple chaotic inflation models with a quadratic potential for some scalar field predict $n_s \simeq 0.96$ and $r \simeq 0.16$. The prediction for $n_s$ is in very good agreement with the present \textit{Planck} data. From the combined current data, the prediction for $r$ is disfavoured at 95\% confidence level \cite{Planck}, and will be further tested with the results from next year. 

When constructing a convincing model of inflation, it is desirable to protect the inflaton potential against unwanted terms which would spoil its flatness. Such terms are otherwise expected due to the large vacuum energy during inflation, which is the so-called $\eta$-problem \cite{Copeland:1994vg}. For chaotic inflation models with a quadratic potential, when embedded in supergravity (SUGRA), it has been shown a long time ago in \cite{Kawasaki:2000yn} that the inflaton potential can be protected by a shift symmetry in the K\"ahler potential. This symmetry can be seen as an approximate symmetry, broken only slightly by a term in the superpotential. On the other hand, the shift symmetry necessarily implies that the inflaton is a gauge singlet. This limits the possible connection to particle physics models.   

As an alternative symmetry solution to the $\eta$-problem, the use of a Heisenberg symmetry has been proposed \cite{Stewart:1994ts,Gaillard:1995az}. Heisenberg symmetry appears for instance in string theory in heterotic orbifold compactifications, where it is a property of the tree-level K\"ahler potential of untwisted matter fields. 
It has been shown in \cite{tribridHeisenberg} that an approximately conserved Heisenberg symmetry solves the $\eta$-problem, and that the associated modulus is stabilized with a large mass by a K\"ahler potential coupling to the field which provides the inflationary vacuum energy via its F-term. 
 It has been applied to tribrid inflation in \cite{tribridHeisenberg,Antusch:2011ei} and to chaotic inflation in \cite{Antusch:2009ty}. One interesting feature of the Heisenberg symmetry is that, in contrast to a shift symmetry, it is not restricted to singlet fields, but can also be used for gauge non-singlet inflaton fields \cite{gnsTribrid}. This  opens up new possibilities for inflation model building.

In this paper, we apply the Heisenberg symmetry to new inflation in the context of SUGRA (cf.\ \cite{Senoguz:2004ky}). We show that this protects the inflaton potential efficiently against corrections which can spoil its flatness. Considering an effective field theory superpotential, we discuss the discrete predictions for the primordial perturbation parameters, which depend only on the order $n$ at which the inflaton appears in the effective superpotential. For $n \ge 5$ the predictions for the spectral index $n_s$ are remarkably close to the best-fit value of the latest \textit{Planck 2013} results (cf.~table \ref{tab:n} and figure \ref{fig:alpha}). While the predictions for $n_s$ are similar to the predictions of chaotic inflation with quadratic potential, SUGRA new inflation with Heisenberg symmetry predicts tiny $r$ which will allow to clearly discriminate it from chaotic inflation with future CMB results.

\section{Superpotentials of new inflation}

We consider an effective theory superpotential of the schematic form
\be\label{eq:Winf}
W_\mathrm{inf} = \kappa S (\Phi^n - M^2)\:,
\ee
with $n\ge 3 $, and where we have used natural units with $M_{\mathrm{Pl}}= (8 \pi G)^{-1/2} = 1$ to keep the notation simple. $S \Phi^n$ is an effective operator, suppressed by $n-2$ powers of the cutoff scale. A cutoff different from the Planck scale can be brought to this form by a redefinition of $\kappa$. 

Superpotentials of this type are frequently encountered in supersymmetric models of particle physics, and can readily be obtained by imposing charges for $S$ and $\Phi$ under the symmetries of the theory. For example, when we impose a $U(1)_\mathrm{R}$ symmetry and a $\mathbb{Z}_n$ symmetry, this would lead directly to the superpotential in eq.~(\ref{eq:Winf}) if we distribute two units of $U(1)_\mathrm{R}$ charge to $S$ and zero to $\Phi$, and one unit of $\mathbb{Z}_n$ charge to $\Phi$ while $S$ is a $\mathbb{Z}_n$-singlet. Additional effective operators allowed by the symmetries which are not written in eq.~(\ref{eq:Winf}) would then be suppressed by high powers of the cutoff scale and can therefore safely be neglected. The above superpotential thus describes a second order phase transition, where the $\mathbb{Z}_n$ symmetry gets spontaneously broken when $\Phi$ takes its vacuum expectation value $\langle \Phi \rangle \simeq M^{2/n}$ in the ground state of the theory. 

A superpotential of the same form as in eq.~(\ref{eq:Winf}) can also describe the spontaneous breaking of a more general symmetry group $G$, when $\Phi^n$ is replaced by a combination of fields such that the product forms a singlet under $G$. For example, one may take $\Phi$ in some complex representation of $G$ (which may for instance be the gauge group of some Grand Unified Theory) and $\bar\Phi$ in the conjugate representation, such that $\bar\Phi \Phi$ forms a singlet. Then eq.~(\ref{eq:Winf}) would take the form $W'_\mathrm{inf} = \kappa S ((\bar\Phi \Phi )^m - M^2)$ (cf.\ \cite{Senoguz:2004ky}). Another example would be that three fields $\Phi_i$, $i \in {1,2,3}$ form a real triplet $\vec\Phi$ under some non-Abelian discrete family symmetry group like $A_4$. Then $\Phi^3$ could for instance be replaced by the $A_4$-invariant product $[\Phi]^3 := \vec\Phi \cdot \vec\Phi \star \vec\Phi$ of three triplets, and eq.~(\ref{eq:Winf}) would take the form $W''_\mathrm{inf} = \kappa S ([\Phi]^{3m} - M^2)$ (cf.\ \cite{Antusch:2008gw}).

\section{K\"ahler potential with Heisenberg symmetry}

We will assume that the K\"ahler potential respects the Heisenberg symmetry with $\Phi$ and a modulus field $T$ (cf.~\cite{Gaillard:1995az}).\footnote{One may in principle also impose a shift symmetry for $\Phi$, however since we want the inflaton to break spontaneously some symmetry, as explained in the previous section, we prefer the use of a  Heisenberg symmetry.} This implies that both fields only appear in the combination
\be
\rho = T + T^* - |\Phi|^2
\ee
in the K\"ahler potential. Besides the Heisenberg symmetry, we will consider a general expansion of fields over the Planck scale, which works very well since all field values will be much below $M_{\mathrm{Pl}}$. Considering terms up to order $|S|^4/M^2_{\mathrm{Pl}}$, we have (in natural units)
\be\label{eq:Kinf}
K_\mathrm{inf} = |S|^2 + \kappa_S |S|^4 + g(\rho)|S|^2 + f(\rho) \:,
\ee
where we have left $g(\rho)$ and $f(\rho)$ here as general functions of $\rho$.  As an explicit example, one may consider $g(\rho) = \kappa_\rho \rho$ and $f(\rho)$ of no-scale form, i.e.\ $f(\rho) = - 3 \ln (\rho)$ as has been analyzed in \cite{tribridHeisenberg} in the context of tribrid inflation. 

\section{The inflationary potential}

From $W_\mathrm{inf}$ and $K_\mathrm{inf}$ one can calculate the scalar potential for the fields $T$, $S$ and $\Phi$. 
As discussed e.g.\ in \cite{tribridHeisenberg}, the K\"ahler potential term $\kappa_S |S|^4$ induces a mass for the scalar component\footnote{We will use the same symbols for the superfields and for their scalar components.} of $S$. Taking $g(\rho) =  \kappa_\rho \rho$ here and in the following as an example, we find (following \cite{tribridHeisenberg}) that the canonically normalized field $\hat S = \sqrt{1+ \kappa_\rho \rho} \,S$ receives a mass of $- 192 \,H^2 \kappa_S$, such that with negative $\kappa_S < -1/192$ the mass of $\hat S$ is positive and larger than the Hubble scale $H$. In the following we will assume that this is the case and that $S$ is stabilized at $S = 0$ during inflation. 

As discussed in \cite{tribridHeisenberg}, one can now change basis to variables $\Phi$ and $\rho$ where the K\"ahler metric becomes diagonal. In our model defined by eqs.~(\ref{eq:Winf}) and (\ref{eq:Kinf}) we obtain for the kinetic terms:\footnote{We will consider negative $f(\rho)$, for instance $f(\rho) = - 3 \ln (\rho)$, such that the kinetic term for $\Phi$ has the right sign.}    
\bea \label{eq:Vphirho}
V(\Phi,\rho) &=& |\kappa|^2 \, | \Phi^n - M^2 |^2 \, \frac{e^{f(\rho)}}{1+ \kappa_\rho \rho} \:, \\ 
{\cal L}_\mathrm{kin} &=& \frac{f''(\rho)}{4} (\partial_\mu \rho)(\partial^\mu \rho) + (- f'(\rho)) (\partial_\mu \Phi )(\partial^\mu \Phi^* )\:.
\eea
During inflation, $\Phi^n \ll M^2$ and thus $|\kappa|^2 \, |\Phi^n - M^2|^2 \approx |\kappa|^2 |M|^4$. For the example $f(\rho) = - 3 \ln (\rho)$, and with negative $\kappa_\rho$,  the potential for $\rho$ has a minimum at $\rho_\mathrm{min} = -\frac{3}{4 \kappa_\rho}$. Including canonical normalization of the $\rho$ field (which will be discussed below), it has a mass of $m_{\hat\rho}^2=24H^2$ (at $\rho = \rho_\mathrm{min}$ where  $H^2 \approx \frac{1}{3} |\kappa|^2 |M|^4  \frac{e^{f(\rho_\mathrm{min})}}{1+ \kappa_\rho \rho_\mathrm{min}}$). We can therefore assume that $\rho$ stabilizes quickly at $\rho = \rho_\mathrm{min}$ before the observable part of inflation starts.

Note that since the potential $V(\Phi,\rho)$ is factorizable in parts depending only on $\Phi$ and only on $\rho$, the minimum for $\rho$ is independent of $\Phi$. $\rho$ is a real scalar field and depends on $\mathrm{Re}(T)$ only. The other component $\mathrm{Im}(T)$ very quickly approaches a constant value in an expanding universe and then decouples from the equations of motion of the other fields (cf.~discussion in \cite{tribridHeisenberg}).  

Setting $\rho = \rho_\mathrm{min}$, the $\rho$-dependent part of the potential just results in a rescaling of the inflaton potential. To calculate the predictions, we will canonically normalize the kinetic terms of the fields by transforming $\rho \to \hat\rho =  \rho\sqrt{f''(\rho_\mathrm{min})/2}$ and
\be
\Phi \to \hat\Phi = \Phi \sqrt{- f'(\rho_\mathrm{min})}\:,
\ee 
and then decompose $\hat\Phi$ in real canonically normalized scalar fields as 
\be
\hat\Phi = \tfrac{1}{\sqrt{2}} (\hat\phi_\mathrm{R}+ \mathrm{i} \hat\phi_\mathrm{I})\:.
\ee

Without loss of generality we can take $M \in \mathbb{R}_+$, absorbing a possible phase by suitable redefinitions of $S$ and $\Phi$. The global minima for $\Phi $ are then at $\Phi = e^{i 2\pi j/n} M^{2/n}$ with $j=1, \dots,n$. When the inflaton has an initial value close to $\Phi = 0$, it will roll towards one of these minima. In what follows, we will consider the trajectory where $\hat\phi_\mathrm{I}=0$ such that the inflaton rolls towards the real minimum. This means we have the following potential for the real field $\hat\phi_\mathrm{R}$:
\be
 V(\hat\phi_\mathrm{R}) = |\kappa|^2 \, \left[ \left(\frac{1}{\sqrt{- f'(\rho_\mathrm{min})} }\, \frac{\hat\phi_\mathrm{R}}{\sqrt{2}}\right)^n - M^2 \right]^2   \frac{e^{f(\rho_\mathrm{min})}}{1+ \kappa_\rho \rho_\mathrm{min}} \:.
\ee 
This expression can be simplified by redefining $\kappa$ and $M$,
\bea
| \kappa |^2 &\to& |\hat\kappa|^2 =  | \kappa |^2\,  \left(- 2 f'(\rho_\mathrm{min}) \right)^{-n/2}  \frac{e^{f(\rho_\mathrm{min})}}{1+ \kappa_\rho \rho_\mathrm{min}} \:,\\
M^2 &\to& \hat M^2 = M^2 \left( - 2 f'(\rho_\mathrm{min})  \right)^{n/2} .
\eea

\begin{figure}[H]
\centering
\includegraphics[width=11cm]{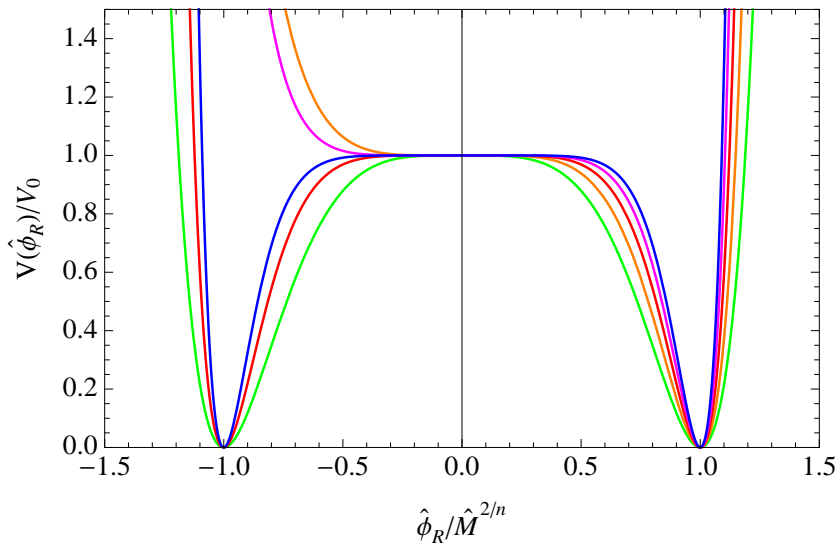}
\caption{Scalar potential $ V(\hat \phi_\mathrm{R})$ for the inflaton field $\hat \phi_\mathrm{R}$, for $n = 4$ (green),  $n = 5$ (orange),  $n = 6$ (red),  $n = 7$ (magenta),  $n = 8$ (blue). It features a plateau for small $\hat \phi_\mathrm{R} \ll \hat M^{2/n}$, which is suitable for slow-roll inflation.}
\label{fig:scalarplot}
\end{figure}

Thus, for obtaining the predictions for the primordial perturbations we can analyze the following simple potential for the real scalar field $\hat\phi_\mathrm{R}$ (with $\hat M \in \mathbb{R}_+$):
\be\label{eq:infpotential}
 V(\hat\phi_\mathrm{R}) = |\hat \kappa|^2 \, \left[ (\hat \phi_\mathrm{R} )^n - \hat{M}^2 \right]^2  .
\ee 
The inflationary potential $V(\hat\phi_\mathrm{R})$ is illustrated in figure \ref{fig:scalarplot} for different values of $n$. It features a plateau for small $\hat\phi_\mathrm{R} \ll \hat M^{2/n}$, which is suitable for inflation as we will discuss in section \ref{predictions}.

Before we turn to the discussion of the predictions for the primordial perturbations, we would like to comment on the question why the inflaton field initially has a small field value. One possible explanation is a period of ``preinflation'' before the observable inflation has started (cf.\ \cite{Senoguz:2004ky}). This can deform the potential such that the inflaton can be temporarily stabilized close to the hilltop. Alternatively, such a stabilizing term can also be provided by a D-flat combination of matter fields $X$, which couple to the inflaton superfield $\Phi$ with a superpotential term proportional to $X^\ell \Phi^m$ (as in tribrid inflation). When the field values of $S$ or of the D-flat combination of matter fields $X$ approach zero, the stabilizing terms disappear and new inflation can start with small inflaton field values.

\section{Predictions for the primordial perturbations}\label{predictions}
With the definition $V_0\equiv|\hat \kappa|^2\,\hat{M}^4$, we can write the inflaton potential of eq.~(\ref{eq:infpotential}) as\footnote{We note that due to the very small inflaton field values, the loop contributions to the potential are very small and can be neglected.}
\be
 V(\hat \phi_\mathrm{R}) =  V_0 \, \left[1-\frac{2(\hat \phi_\mathrm{R})^n}{\hat{M^2}} + \frac{(\hat \phi_\mathrm{R})^{2n}}{\hat{M^4}}\right]  \:.
 \label{eq:Vscalar}
\ee 
Since inflation will happen for small field values, the last term in eq.~(\ref{eq:Vscalar}) can be neglected. The inflaton potential reduces to a simple form, from which the predictions for the primordial perturbations can readily be derived in the slow roll approximation. The slow roll parameters $\epsilon$, $\eta$ and $\xi^2$ are given as functions of $\hat \phi_\mathrm{R}$ by
\bea
\epsilon(\hat \phi_\mathrm{R}) &=& \frac{1}{2}\left(\frac{V'}{V} \right)^2 \approx \frac{1}{2}\left(2n\frac{(\hat \phi_\mathrm{R} )^{n-1}}{\hat M^2} \right)^2 \:,\\
\label{eq:Epsilon}
\eta(\hat \phi_\mathrm{R}) &=& \frac{V''}{V} \approx -2n\,(n-1)\,\frac{(\hat \phi_\mathrm{R} )^{n-2}}{\hat M^2}\:,\\
\label{eq:Eta}
\xi^2(\hat \phi_\mathrm{R}) &=& \frac{V'V'''}{V^2} \approx 4n^2\,(n-1)\,(n-2)\,\frac{(\hat \phi_\mathrm{R} )^{2n-4}}{\hat M^4} \:.
\label{eq:Xi}
\eea
Inflation ends when one of the slow-roll conditions are violated. Here, the relevant condition is $|\eta(\hat \phi^f_\mathrm{R})| \approx 1$ or equivalently 
\be
\hat \phi^f_\mathrm{R} \approx \left(\frac{\hat{M}^2}{2n\,(n-1)}\right)^{\frac{1}{n-2}} \:.
\label{eq:Phiend}
\ee
To calculate the predictions for the inflationary observables we need to know the inflation field value when the relevant perturbations cross the horizon, which is typically at about $N_i = 50$ to $60$ $e$-folds before the end of inflation. Calling this field value $\hat \phi^i_\mathrm{R}$ the number of $e$-folds $N_i$ is given by 
\be
N_i = \int_{\hat \phi^f_\mathrm{R}}^{\hat \phi^i_\mathrm{R}} \frac{V\,d\hat \phi_\mathrm{R}}{V'} 
\approx \frac{\hat{M}^2}{2n\,(n-2)(\hat \phi^i_\mathrm{R})^{n-2}}-\frac{n-1}{n-2}\: .
\label{eq:N}
\ee
This equation can be solved for $\hat \phi^i_\mathrm{R}$ as a function of $N_i$ and plugged into the slow roll parameters to calculate the parameters for the primordial perturbations. 
This yields the following predictions for the spectral index $n_s$, the running of the spectral index $\alpha_s = \frac{dn_s}{d \ln k} $ and the tensor-to-scalar ratio $r$:
\bea
n_s &\approx& 1+ 2\eta \approx 1-\frac{2\,(n-1)}{(n-2)\,N_i+(n-1)} \:,\\
\alpha_s &\approx& 16\epsilon\eta - 24\epsilon^2 -2\xi^2 \approx -2\xi^2 \approx -2\,\frac{(n-1)(n-2)}{\left[(n-2)\,N_i+(n-1) \right]^2}\:,\\
r &\approx& 16\,\epsilon \approx \frac{32\,n^2}{\hat{M}^4}\left[\frac{\hat{M}^2}{2n\,\left[(n-2)\,N_i+(n-1)\right]}\right]^{\frac{2n-2}{n-2}}\; .
\label{eq:n_s}
\eea
The predictions for $n_s$ and $\alpha_s$ depend only on $N_i$,\footnote{The precise value of $N_i$ depends on the later evolution of the universe and requires for instance the calculation of the reheating and preheating phase. In an explicit model, where the couplings of the inflaton field are known, it can in principle be computed. Here we will present the results for values of $N_i$ between $50$ and $60$, as also done e.g.\ in \cite{Planck}. } while the prediction for $r$ depends also on $\hat{M}$. As usual for small field inflation, $r$ is tiny ($< {\cal O}(10^{-7})$), below observational possibilities. The predictions for $n_s$ and $\alpha_s$ are given in table \ref{tab:n} for $N_i = 60$ and illustrated in figure \ref{fig:alpha} with $N_i$ ranging from $50$ to $60$. 
For comparison, the \textit{Planck 2013} results (at 1$\sigma$) are $n_s = 0.9603 \pm 0.0073$ and $\alpha_s = - 0.013 \pm 0.009$ \cite{Planck}. The slight hint in the \textit{Planck 2013} results for non-zero $\alpha_s$ is not at a statistically significant level. The model prediction for $n_s$ is in good agreement for $n=4$ and in excellent agreement for larger $n$.

\begin{table}
\centering
\begin{tabular}{c|c|c|c|c|c|c}
	\textbf{n} & \textbf{3} & \textbf{4} & \textbf{5} & \textbf{6} & \textbf{7} & \textbf{8} \\
	\hline\hline
	$n_s$ & 0.935 & 0.951 & 0.957 & 0.959 & 0.961 & 0.962 \\
	$\alpha_s / 10^{-3}$ & -1.041 & -0.793 & -0.709 & -0.666 & -0.641 & -0.624 \\

	\hline
\end{tabular}
\caption{Predictions for the spectral index $n_s$ and its running $\alpha_s = \frac{dn_s}{d \ln k}$ for different $n$, calculated with $N_i=60$.}
\label{tab:n}
\end{table}

\begin{figure}
\centering
\includegraphics[width=11cm]{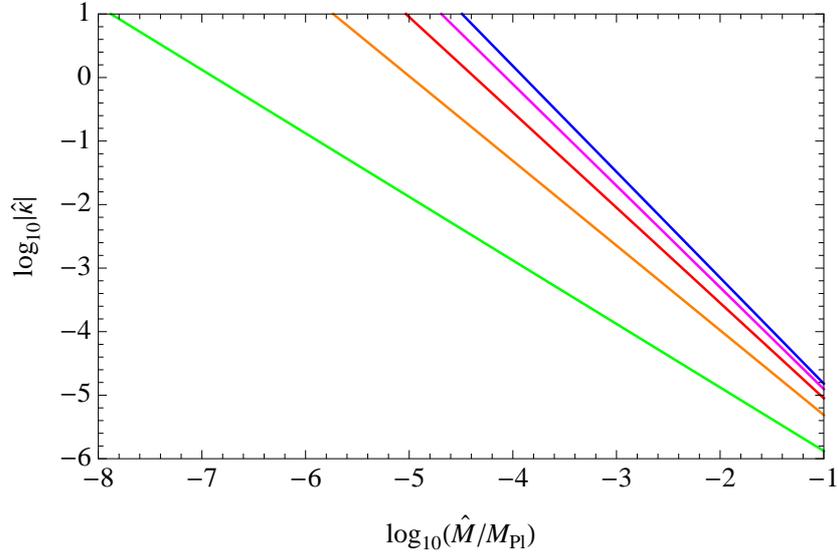}
\caption{Relation between the parameters $|\hat{\kappa}|$ and $\hat{M}/M_{Pl}$ of the inflaton potential (cf.\ eq.~(\ref{eq:infpotential})) for $n = 4$ (green),  $n = 5$ (orange),  $n = 6$ (red),  $n = 7$ (magenta),  $n = 8$ (blue), calculated from eq.~(\ref{eq:As}) with $A_s = 2.215 \times 10^{-9}$ \cite{Ade:2013zuv}.
}
\label{fig:kappa}
\end{figure}

The amplitude $A_s$ of the scalar perturbations is given as
\be
A_s \approx \frac{V_0}{24\pi^2\epsilon}\approx\frac{|\hat\kappa|^2 \hat{M}^8}{48\pi^2n^2}\,\left[\frac{\hat{M}^2}{2n\left[(n-2)\,N_i+(n-1)\right]}\right]^{-\frac{2n-2}{n-2}}\:.
\label{eq:As}
\ee
Using the best-fit value measured by \textit{Planck}, $A_s = 2.215 \times 10^{-9}$ \cite{Ade:2013zuv}, one can obtain $\hat{M}$ as a function of $|\hat \kappa|$ as shown in figure \ref{fig:kappa} .

\section{Summary}

We have proposed a realisation of ``new inflation'' in supergravity, where the flatness of the inflaton potential is protected by a Heisenberg symmetry in the K\"ahler potential. Inflation can be associated with a particle physics phase transition, with the inflaton being a (D-flat) direction of Higgs fields which break some symmetry at high energies, e.g.\ of GUT Higgs fields or of Higgs fields for flavour symmetry breaking. This is possible since compared, e.g., to shift symmetry, which is usually used to protect a flat inflaton potential, the Heisenberg symmetry is compatible with a (gauge) non-singlet inflaton field. 

With field values much below the Planck scale $M_\mathrm{Pl}$, we have considered an expansion of the K\"ahler potential and superpotential in fields over $M_\mathrm{Pl}$. The modulus field associated with the Heisenberg symmetry is stabilized at its minimum during inflation, with a mass much larger than the Hubble scale, and the potential has trajectories where inflation can proceed along a very flat plateau. 
Suitable initial conditions close to the top of the hill can be justified by a stage of ``preinflation'', before the observable part of inflation starts, when for instance the singlet driving field $S$ or a D-flat combination of matter fields (as in tribrid inflation) has a non-vanishing vacuum expectation value. This can temporarily stabilize the inflaton field close to the hilltop and can explain the small inflaton field values at the beginning of inflation.

\begin{figure}
\centering
\includegraphics[width=11.5cm]{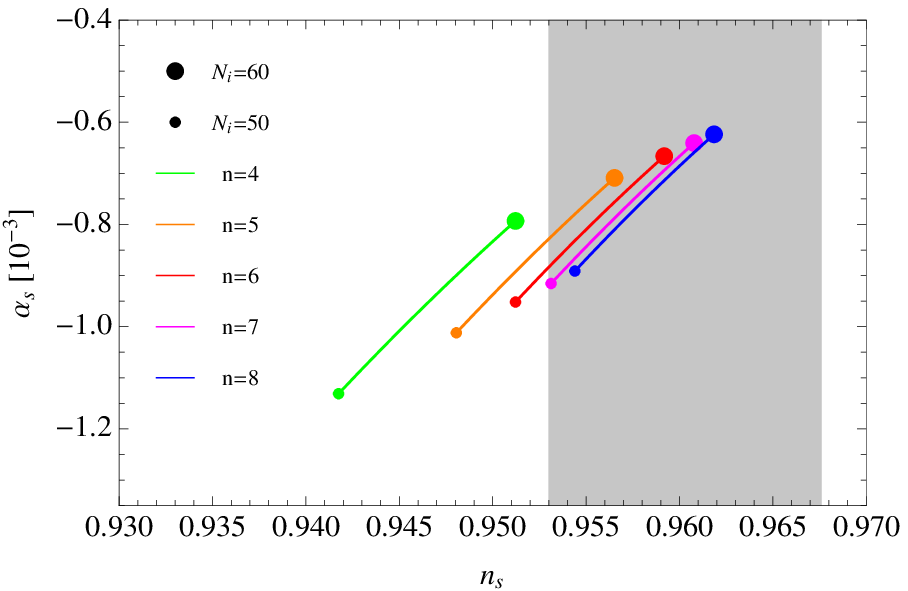}
\caption{Predictions for the spectral index $n_s$ and for its running $\alpha_s = \frac{dn_s}{d \ln k}$, for different $n$ and with $N_i$ ranging from $50$ to $60$. The grey region shows the \textit{Planck 2013} $1\sigma$ range for $n_s$.}
\label{fig:alpha}
\end{figure}

In contrast to conventional new inflation models in SUGRA where the predictions depend on unknown parameters of the K\"ahler potential, the model with Heisenberg symmetry makes discrete predictions for the parameters describing the primordial perturbations, e.g.\ for the spectral index $n_s$ and the running of the spectral index $\alpha_s$ (cf.~table \ref{tab:n} and figure \ref{fig:alpha}), which depend only on the order $n$ at which the inflaton appears in the effective superpotential. The predictions for $n_s$ for $n\ge 5$ are close to the best-fit value of the latest \textit{Planck 2013} results.

\section*{Acknowledgements}
This project was supported by the Swiss National Science Foundation. We thank David Nolde and Stefano Orani for useful discussions.


\end{document}